\documentclass[aps,prb,twocolumn,showpacs,preprintnumbers,amsmath,amssymb,superscriptaddress]{revtex4}

\usepackage{graphicx}
\usepackage{dcolumn}
\usepackage{bm}

\newcommand{\ket}[1]{|#1\rangle}

\begin{document}

\title{Shot noise and tunnel magnetoresistance in multilevel
quantum dots: Effects of cotunneling}

\author{I. Weymann}
\email{weymann@amu.edu.pl} \affiliation{Department of Physics,
Adam Mickiewicz University, 61-614 Pozna\'n, Poland}

\author{J. Barna\'s}
\affiliation{Department of Physics, Adam Mickiewicz University,
61-614 Pozna\'n, Poland} \affiliation{Institute of Molecular
Physics, Polish Academy of Sciences, 60-179 Pozna\'n, Poland}

\date{\today}

\begin{abstract}
Spin-dependent transport through a multilevel quantum dot weakly
coupled to ferromagnetic leads is analyzed theoretically by means
of the real-time diagrammatic technique. Both the sequential and
cotunneling processes are taken into account, which makes the
results on tunnel magnetoresistance (TMR) and shot noise
applicable in the whole range of relevant bias and gate voltages.
Suppression of the TMR due to inelastic cotunneling and
super-Poissonian shot noise have been found in some of the Coulomb
blockade regions. Furthermore, in the Coulomb blockade regime
there is an additional contribution to the noise due to bunching
of cotunneling processes involving the spin-majority electrons. On
the other hand, in the sequential tunneling regime TMR oscillates
with the bias voltage, while the current noise is generally
sub-Poissonian.
\end{abstract}


\maketitle

{\it Introduction:} Transport properties of quantum dots coupled
to ferromagnetic leads are currently a subject of extensive
experimental and theoretical studies.
\cite{maekawa02,zutic04,yakushijiPR07,
rudzinski01,braun04,cottet04,weymannPRB05,weymannPRBBR05,kondo,
ralph02,heersche06,fertAPL06,hamayaAPL07a,hamayaAPL07b,hamaya07}
This interest is stimulated by expected applications in
spintronics and quantum computing.
\cite{maekawa02,zutic04,yakushijiPR07} When a quantum dot is
coupled to two ferromagnetic leads, its transport properties
depend on the magnetic configuration of the system. This is the
so-called tunnel magnetoresistance (TMR) effect, which is
characterized by the ratio ${\rm TMR} = (I_{\rm P} - I_{\rm
AP})/I_{\rm AP}$, where $I_{\rm P}$ ($I_{\rm AP}$) is the current
flowing through the system in the parallel (antiparallel)
configuration. \cite{julliere75,rudzinski01,weymannPRB05} When the
coupling between the dot and leads is strong, the Kondo physics
emerges for $T\lesssim T_{\rm K}$, where $T_{\rm K}$ is the Kondo
temperature. \cite{kondo} In turn, single-electron charging in the
weak coupling regime leads to the Coulomb blockade phenomena.
\cite{rudzinski01,braun04} Sequential (first order) transport is
exponentially suppressed in the blockade regime. The current flows
then due to cotunneling (second order) processes involving
correlated tunneling through virtual states of the dot, whereas
outside the blockade regime transport is dominated by sequential
tunneling processes. \cite{cotunneling} So far fully systematic
considerations (taking into account both sequential and
cotunneling processes) of spin-dependent transport through a
quantum dot in the weak coupling regime have been restricted
mainly to single-level quantum dots. \cite{weymannPRB05} In real
systems, however, usually more than one energy level participate
in transport, leading to more complex and interesting transport
characteristics. \cite{belzigPRB04,weymannJPCM07} It has been
shown recently that the Fano factor in the Coulomb blockade regime
calculated in the first order approximation is larger than unity.
\cite{weymannJPCM07} Moreover, the TMR was found then to be
independent of the gate voltage. In this paper we extend the
theoretical studies by including the cotunneling processes, and
show that the shot noise in the Coulomb blockade regime can be
super-Poissonian, although the Fano factor is significantly
reduced by the cotunneling processes. Apart from this, we show
that the TMR in the blockade regime is considerably modified by
cotunneling processes, and can be either enhanced or reduced in
comparison to that in the first order approximation, depending on
the transport regime. Our considerations are based on the
real-time diagrammatic technique
\cite{weymannPRB05,diagrams,thielmann} which, after taking into
account both the first and second-order contributions, allows us
to analyze transport in the {\it full} weak coupling regime, i.e.
in the cotunneling, cotunneling-assisted sequential, and
sequential tunneling regimes. Furthermore, to calculate the shot
noise in the cotunneling regime, we include the non-Markovian
effects, \cite{thielmann} which were neglected in previous
considerations.\cite{sukhorukovPRB01}

{\it Model:} We consider a two-level quantum dot weakly coupled to
external ferromagnetic leads whose magnetic moments are either
parallel or antiparallel. The Hamiltonian of the system reads,
$\hat{H}=\hat{H}_{\rm L} + \hat{H}_{\rm R} + \hat{H}_{\rm D} +
\hat{H}_{\rm T}$. The first two terms describe noninteracting
itinerant electrons in the leads, $\hat{H}_r=\sum_{{\mathbf
k}\sigma} \varepsilon_{r{\mathbf k}\sigma} c^{\dagger}_{r{\mathbf
k}\sigma} c_{r{\mathbf k}\sigma}$ for the left ($r={\rm L}$) and
right ($r={\rm R}$) leads, where $c^{\dagger}_{r{\mathbf
k}\sigma}$ ($c_{r{\mathbf k}\sigma}$) creates (annihilates) an
electron with the wave vector ${\mathbf k}$ and spin $\sigma$ in
the lead $r$, and $\varepsilon_{r{\mathbf k}\sigma}$ is the
corresponding dispersion relation. The quantum dot is described by
\begin{equation}\label{Eq:Hamiltonian}
  \hat{H}_{\rm D} =\sum_{j\sigma} \varepsilon_{j} n_{j\sigma}
  + U \sum_j n_{j\uparrow} n_{j\downarrow}
  + U^\prime \sum_{\sigma\sigma^\prime}
  n_{1\sigma}n_{2\sigma^\prime} \,,
\end{equation}
where $n_{j\sigma} = d^{\dagger}_{j\sigma}d_{j\sigma}$ and
$d^{\dagger}_{j\sigma}$ ($d_{j\sigma}$) is the creation
(annihilation) operator of an electron with spin $\sigma$ in the
$j$th level ($j=1,2$), $\varepsilon_{j}$ is the corresponding
single-particle energy, and $U$ ($U^\prime$) is the on-level
(inter-level) Coulomb repulsion parameter. The tunnel Hamiltonian,
$\hat{H}_{\rm T}$, takes the form: $\hat{H}_{\rm T}=\sum_{r=\rm
L,R} \sum_{{\mathbf k}j\sigma} (t_{rj}c^{\dagger}_{r {\mathbf
k}\sigma}d_{j\sigma}+ t_{rj}^\star d^\dagger_{j\sigma} c_{r
{\mathbf k}\sigma} )$, where $t_{rj}$ is the relevant tunneling
matrix element. Coupling of the $j$th level to the spin-majority
(spin-minority) electron band of the lead $r$ is described by
$\Gamma_{rj}^{+(-)}=2\pi |t_{rj}|^2\rho_r^{+(-)}= \Gamma_{rj}(1\pm
p_{r})$, where $\Gamma_{rj}= (\Gamma_{rj}^{+}
+\Gamma_{rj}^{-})/2$, while $\rho_r^{+(-)}$ and $p_r$ are the
spin-dependent density of  states and spin polarization in the
lead $r$, respectively. In the following we assume $\Gamma_{rj}
\equiv \Gamma/2$ and $p_{\rm L} = p_{\rm R}\equiv p$.

{\it Method:} In order to calculate the spin-polarized transport
through a two-level quantum dot in the sequential and cotunneling
regimes, we employ the real-time diagrammatic technique,
\cite{weymannPRB05,diagrams,thielmann} which consists in a
systematic expansion of the quantum dot (reduced) density matrix
and the current operator with respect to the dot-lead coupling
strength $\Gamma$. The current operator $\hat{I}$ is defined as
$\hat{I}=(\hat{I}_{\rm R} - \hat{I}_{\rm L})/2$, with
$\hat{I}_r=-i(e/\hbar) \sum_{{\mathbf k}\sigma}
\sum_{j}(t_{rj}c^{\dagger}_{r {\mathbf k}\sigma} d_{j\sigma}-
t_{rj}^\star d^\dagger_{j\sigma} c_{r {\mathbf k}\sigma})$ being
the current flowing from the dot to the lead $r$. Time evolution
of the reduced density matrix can be visualized as a sequence of
irreducible self-energy blocks, $W_{\chi \chi^\prime}$, on the
Keldysh contour. The matrix elements $W_{\chi \chi^\prime}$
describe transitions between the many-body states $\ket{\chi}$ and
$\ket{\chi^\prime}$ of the two-level dot.\cite{weymannJPCM07} The
full propagation of the dot density matrix is given by the Dyson
equation, which is further transformed into a general kinetic
equation for the elements of the reduced density matrix. With the
aid of the matrix notation introduced in Ref.
~[\onlinecite{thielmann}], all the quantities of interest can be
defined in terms of the following self-energy matrices:
$\mathbf{W}$, $\mathbf{W}^{\rm I}$, $\mathbf{W}^{\rm II}$,
$\partial\mathbf{W}$, and $\partial\mathbf{W}^{\rm I}$. The matrix
$\mathbf{W}^{\rm I (II)}$ is the self-energy matrix with one {\it
internal} vertex (two internal vertices) resulting from the
expansion of the tunneling Hamiltonian replaced by the current
operator, while $\partial\mathbf{W}$ and $\partial\mathbf{W}^{\rm
I}$ are partial derivatives of $\mathbf{W}$ and $\mathbf{W}^{\rm
I}$ with respect to the convergence factor of the Laplace
transform. \cite{thielmann} Using the above matrices, the
stationary occupation probabilities can be found from,
$(\mathbf{\tilde{W}}\mathbf{p}^{\rm st})_{\chi} =
\Gamma\delta_{\chi\chi_0}$\,,
where $\mathbf{p}^{\rm st}$ is the vector containing probabilities
and the matrix $\mathbf{\tilde{W}}$ is given by $\mathbf{W}$ with
one arbitrary row $\chi_0$ replaced by $(\Gamma,\dots,\Gamma)$ due
to the normalization, $\sum_{\chi}p_{\chi}^{\rm st}=1$. The
current flowing through the system can be then found from
\begin{equation}\label{Eq:current}
  I=\frac{e}{2\hbar}{\rm Tr}\{\mathbf{W}^{\rm I}\mathbf{p}^{\rm st}\} \,.
\end{equation}
Finally, the zero-frequency current noise, $S=2\int_{-\infty}^0 dt
(\langle \hat{I}(t)\hat{I}(0)+\hat{I}(0)\hat{I}(t)\rangle-2
\langle \hat{I}\rangle^2 )$, is given by \cite{thielmann}
\begin{equation}\label{Eq:noise}
  S = \frac{e^2}{\hbar}{\rm Tr}\left\{\left[
  \mathbf{W}^{\rm II}+
  \mathbf{W}^{\rm I} \left(
  \mathbf{P}\mathbf{W}^{\rm I} +
  \mathbf{p}^{\rm st} \otimes \mathbf{e}^{\rm T}\partial\mathbf{W}^{\rm
  I}\right)\right]
  \mathbf{p}^{\rm st}
  \right\} \,,
\end{equation}
where the object $\mathbf{P}$ is calculated from:
$\mathbf{\tilde{W}}\mathbf{P} = \mathbf{\tilde{1}}(\mathbf{p}^{\rm
st}\otimes \mathbf{e}^{\rm T} - \mathbf{1} -
\partial\mathbf{W}\mathbf{p}^{\rm st}\otimes \mathbf{e}^{\rm T})$,
with $\mathbf{\tilde{1}}$ being the unit vector with row $\chi_0$
set to zero, and $\mathbf{e}^{\rm T} = (1,\dots,1)$.
\cite{thielmann}

To calculate the transport properties order by order in tunneling
processes, we expand the self-energy matrices, $\mathbf{W}^{\rm
(I,II)} = \mathbf{W}^{\rm (I,II)(1)} +\mathbf{W}^{\rm
(I,II)(2)}+\dots$, the dot occupations, $\mathbf{p}^{\rm st} =
\mathbf{p}^{\rm st (0)} + \mathbf{p}^{\rm st (1)}+\dots$, and,
$\mathbf{P} = \mathbf{P}^{(-1)} + \mathbf{P}^{(0)}+\dots$,
respectively. The self-energies can be calculated using the
corresponding diagrammatic rules. \cite{thielmann,weymannPRB05}
The first order of expansion corresponds to the sequential
tunneling, whereas the second one to cotunneling. Thus, by taking
into account all the first and second-order contributions, we are
able to resolve transport properties in the full range of the bias
and gate voltages.

\begin{figure*}[t]
  \includegraphics[height=7cm]{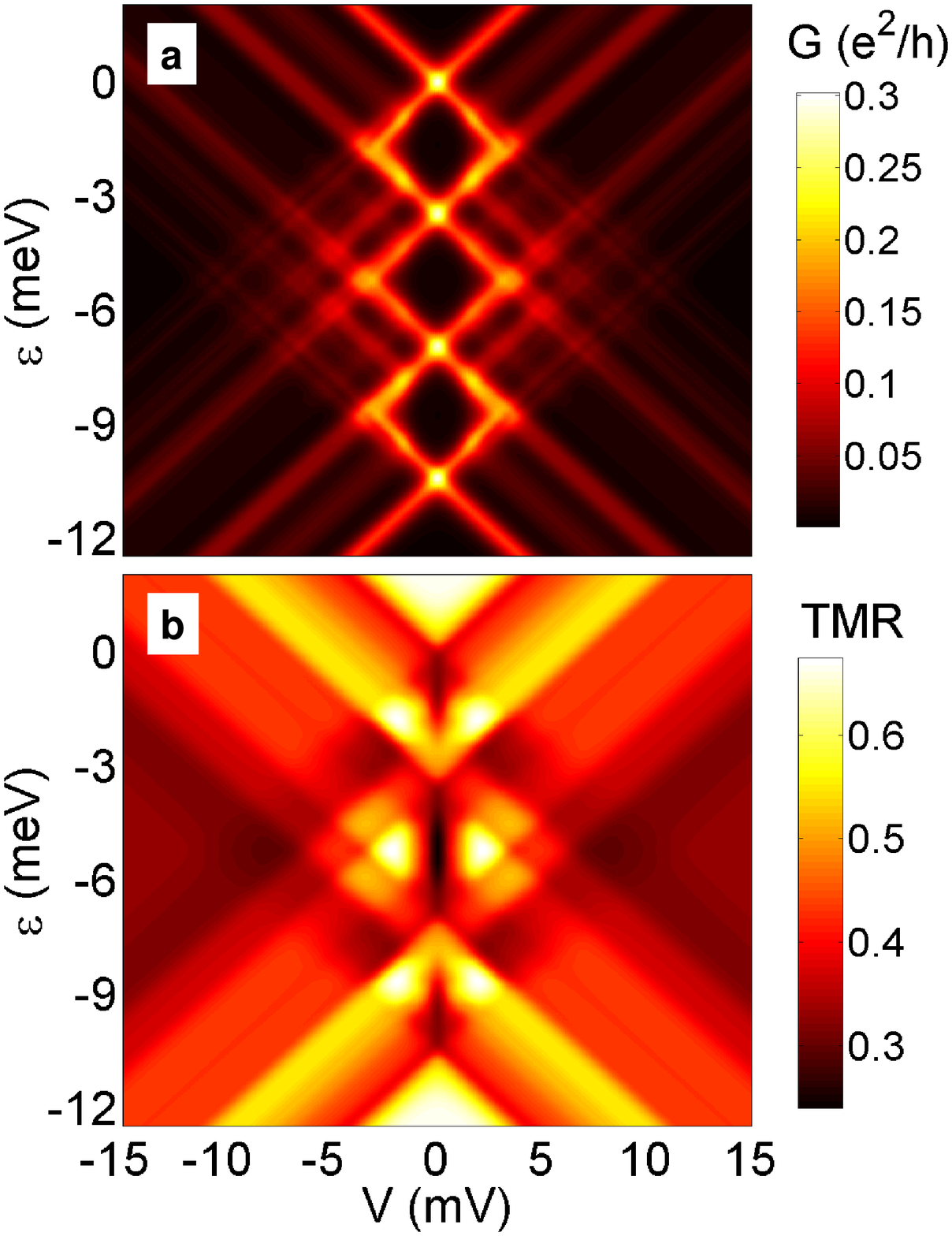}
  \hspace{0.4cm}
  \includegraphics[height=7cm]{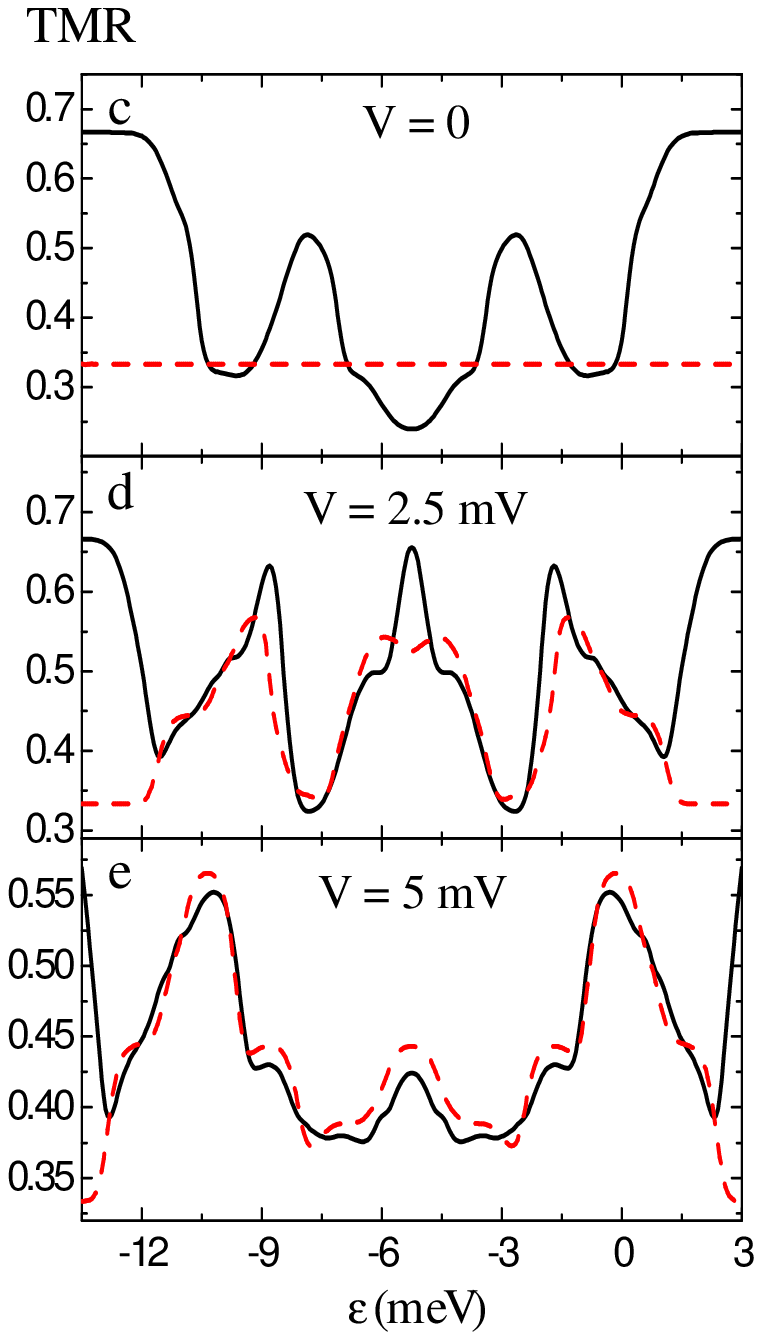}
  \hspace{0.25cm}
  \includegraphics[height=7cm]{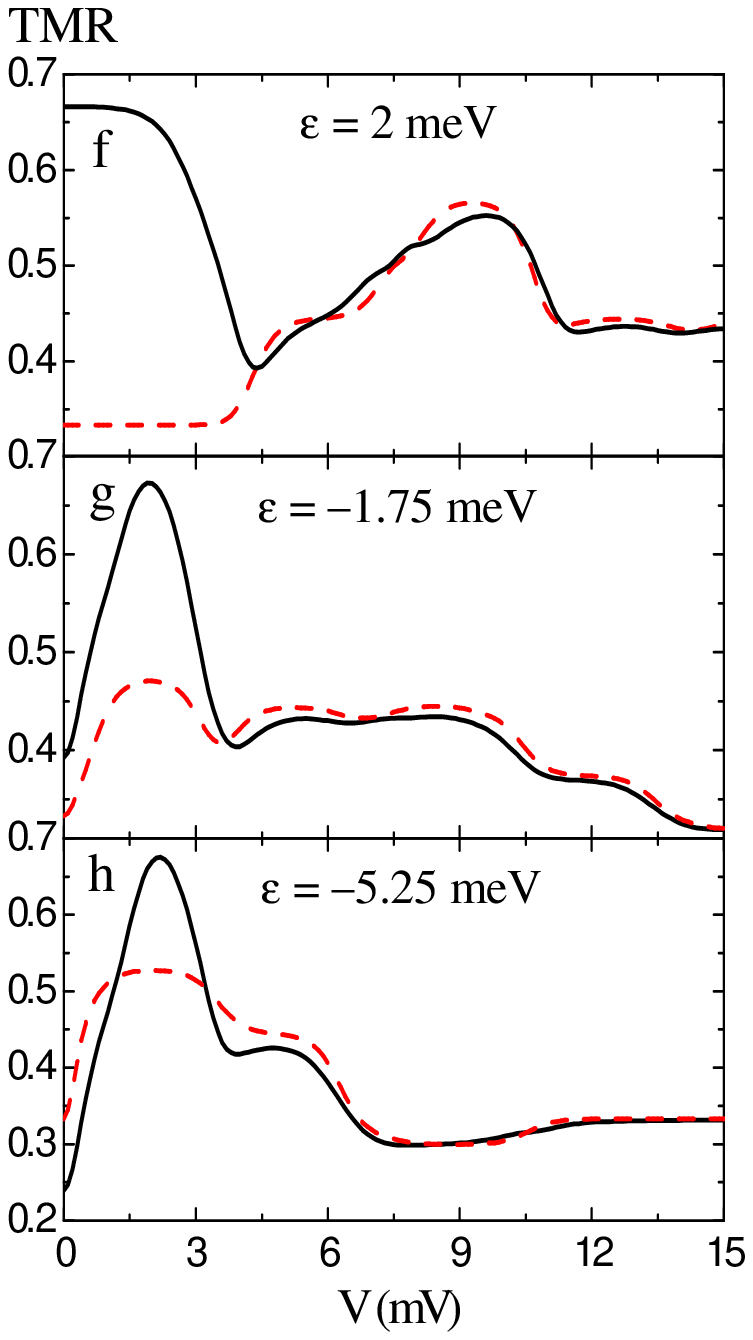}
  \caption{\label{Fig:1} (Color online)
  The differential conductance $G$ in the parallel configuration (a)
  and the TMR (b) as a function
  of the bias voltage $V$ and level position $\varepsilon =
  \varepsilon_1$. Parts (c)-(e) show the TMR as a function
  of $\varepsilon$ for several values of $V$,
  whereas parts (f)-(h) display the bias
  voltage dependence of TMR for several values of $\varepsilon$.
  The dashed lines in (c)-(h) show the first-order contributions.
  The parameters are:
  $k_{\rm B} T=0.15$ meV, $\varepsilon_2-\varepsilon_1=1.5$ meV,
  $U=5$ meV, $U^\prime=2$ meV, $\Gamma=0.1$ meV,
  and $p=0.5$.}
\end{figure*}

{\it Results on TMR:} The TMR as a function of the bias voltage
and position of the dot levels is shown in Fig.~\ref{Fig:1}(b). To
facilitate the identification of different transport regimes, we
present in Fig.~\ref{Fig:1}(a) the density plot of the
differential conductance $G$ in the parallel configuration
(differential conductance in the antiparallel configuration is
qualitatively similar\cite{weymannJPCM07}). Since position of the
dot levels can be shifted with a gate voltage,
Figs.~\ref{Fig:1}(a) and (b) can be viewed as a bias and gate
voltage dependence of TMR and $G$, respectively. By sweeping the
gate voltage in the linear response regime, the number of
electrons in the dot can be changed successively. More precisely,
this happens when: $\varepsilon\equiv\varepsilon_1=0$,
$\varepsilon=-(\delta\varepsilon+U^\prime)$,
$\varepsilon=-(U+U^\prime)$, and
$\varepsilon=-(\delta\varepsilon+U+2U^\prime)$, where
$\delta\varepsilon = \varepsilon_2-\varepsilon_1$ is the level
spacing. Thus, for $\varepsilon \gtrsim 0$
[$-(\delta\varepsilon+U+2U^\prime) \gtrsim \varepsilon$], the dot
is empty (fully occupied). When  $0 \gtrsim \varepsilon \gtrsim
-(\delta\varepsilon+U^\prime)$ [$-(U+U^\prime) \gtrsim \varepsilon
\gtrsim -(\delta\varepsilon+U+2U^\prime)$], there is a single
electron (three electrons) in the dot. On the other hand, for
$-(\delta\varepsilon+U^\prime) \gtrsim \varepsilon \gtrsim
-(U+U^\prime)$, the dot is occupied by two electrons, one on each
orbital level.

In the case of empty (fully occupied) dot in the Coulomb blockade
regime the current flows only due to elastic non-spin-flip
cotunneling processes. Such processes are fully coherent and do
not affect the charge and spin state of the dot. As a result, the
system behaves as a single ferromagnetic tunnel junction, yielding
the TMR given by the Julliere formula,
\cite{weymannPRB05,julliere75} ${\rm TMR} = 2p^2/(1-p^2)$.
However, in the other blockade regions both the non-spin-flip and
spin-flip cotunneling processes are allowed, leading to the
suppression of the TMR. This can be seen in Fig.~\ref{Fig:1}(c),
where we plot the gate voltage dependence of the TMR in the linear
response regime. When there is a single electron on one of the two
orbital levels, the TMR is decreased. This is due to the spin-flip
cotunneling processes which provide a channel for spin relaxation,
decreasing the difference between conductance in the parallel and
antiparallel configurations and thus reducing the TMR.
\cite{weymannPRB05} On the other hand, in the regime when each dot
level is singly occupied (doubly occupied dot), the amount of
spin-flip cotunneling is increased and the suppression of TMR is
even more pronounced. However, when the bias voltage is increased,
the central minimum in TMR [Fig.~\ref{Fig:1}(c)] transforms into a
local maximum due to a nonequilibrium spin accumulation in the
dot, see Fig.~\ref{Fig:1}(d) and (e).

The bias voltage dependence of the TMR for several values of the
level position is displayed in Fig.~\ref{Fig:1}(f)-(h). When the
dot is empty in equilibrium ($\varepsilon=2$ meV), the TMR in the
cotunneling regime is given by the Julliere value. However, once
the bias voltage reaches the threshold voltage ($V\approx 3$ mV),
the sequential processes are allowed and TMR drops, see
Fig.~\ref{Fig:1}(f). If the ground state is singly occupied
($\varepsilon=-1.75$ meV), a nonequilibrium spin accumulation in
doublet states ($p^{\rm st }_{\ket{\uparrow 0}} \neq p^{\rm st
}_{\ket{\downarrow 0}}$) is built up with increasing bias voltage.
\cite{weymannPRBBR05} This leads to an enhanced TMR, which again
starts to drop around the threshold for sequential tunneling, see
Fig.~\ref{Fig:1}(g). Finally, in Fig.~\ref{Fig:1}(h) we show the
bias dependence of TMR for the case when the dot is doubly
occupied in equilibrium ($\varepsilon=-5.25$ meV). Now, the spin
accumulation in triplet states ($p^{\rm st }_
{\ket{\uparrow\uparrow}} \neq p^{\rm st }_
{\ket{\downarrow\downarrow}}$) gives rise to an increase in TMR.
\cite{weymannEPL06} We also note that in the transport regime
where the sequential tunneling is allowed,  more and more charge
states become active in transport with increasing the bias
voltage. This gives rise to step-like $I-V$ characteristics and
the oscillatory-like behavior of the TMR, see
Fig.~\ref{Fig:1}(f)-(h).

\begin{figure*}[t]
  \includegraphics[height=7cm]{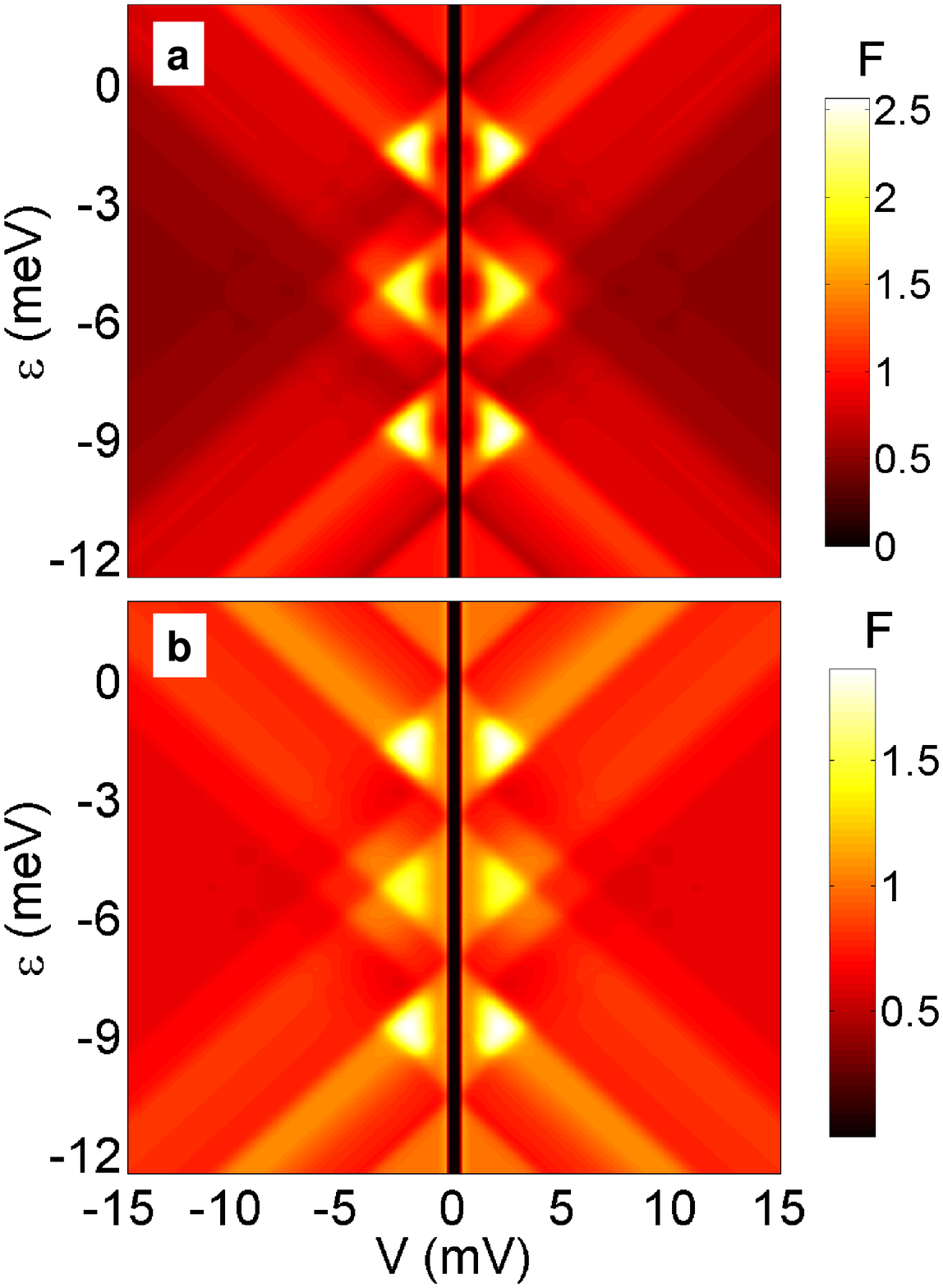}
  \hspace{0.4cm}
  \includegraphics[height=7cm]{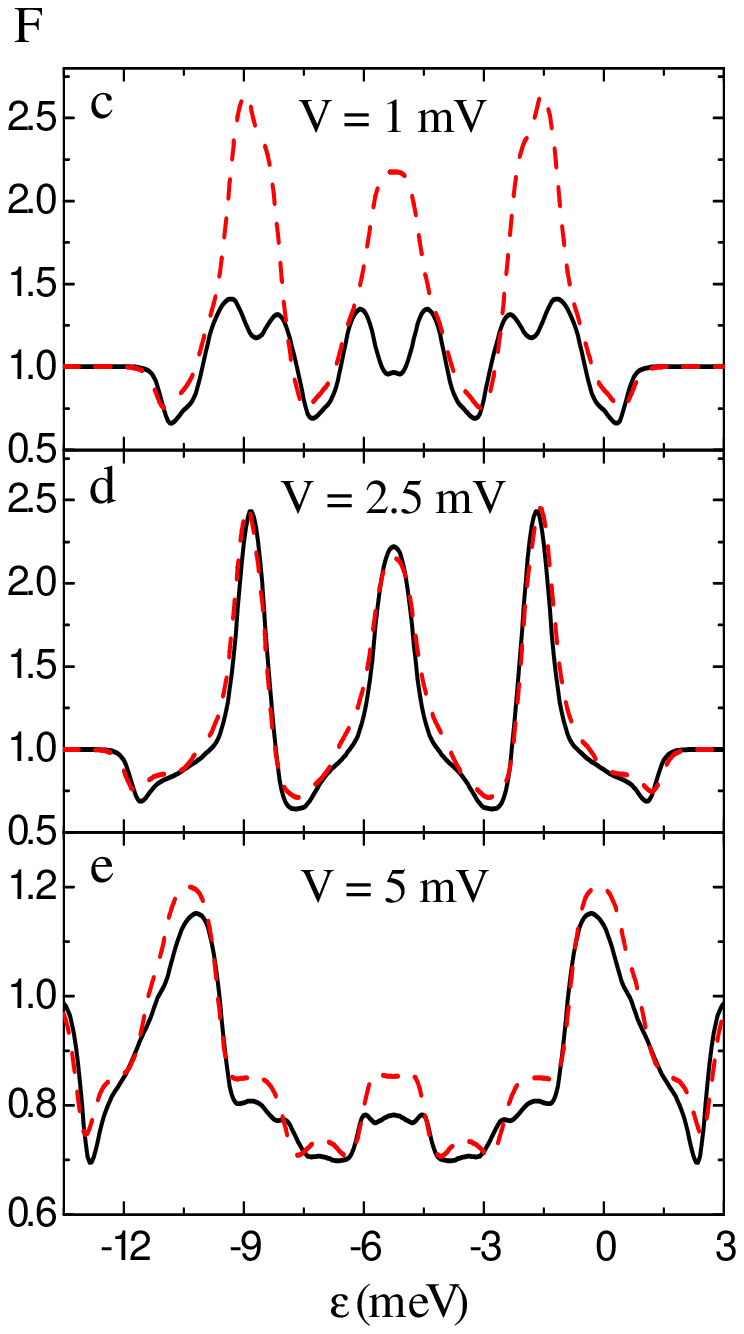}
  \hspace{0.25cm}
  \includegraphics[height=7cm]{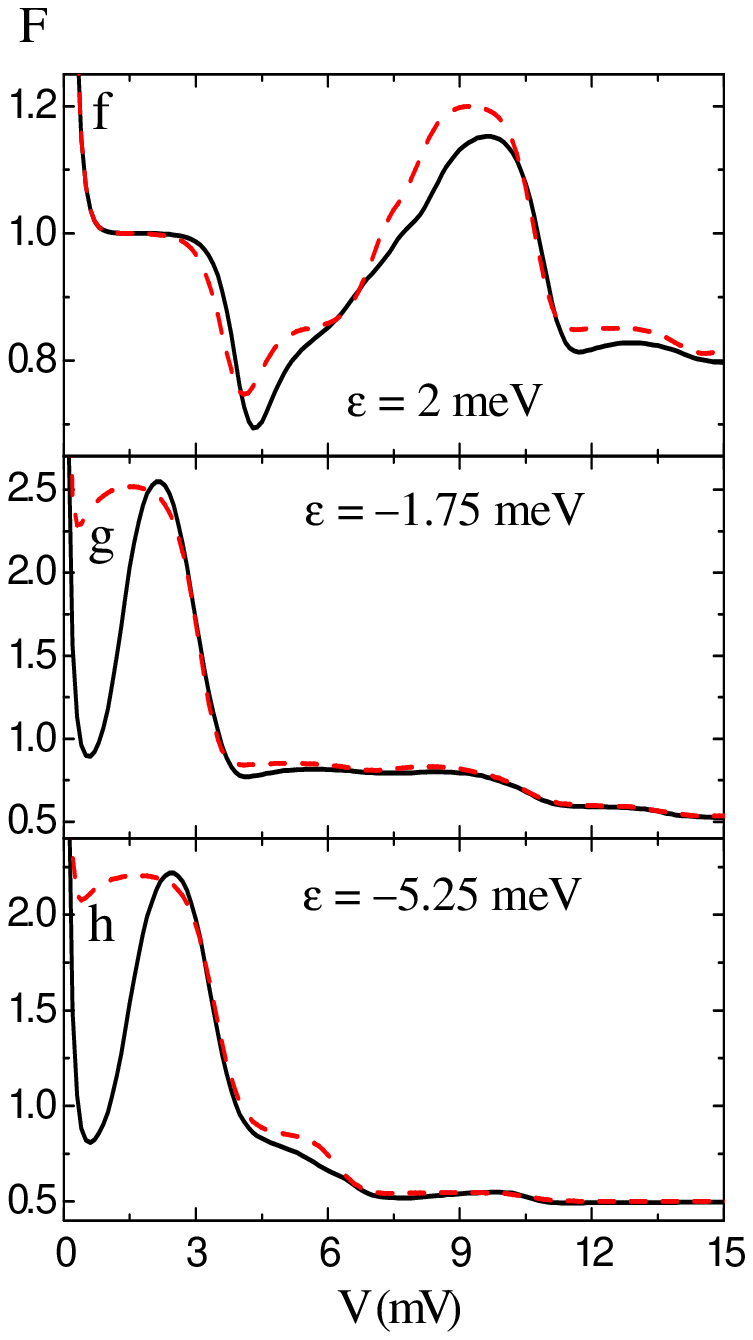}
  \caption{\label{Fig:2} (Color online)
  The Fano factor $F$ as a function of the bias voltage and level position
  for the parallel (a) and antiparallel (b)
  magnetic configurations.
  Parts (c)-(h) show $F$ in the parallel configuration
  as a function of the level position [(c)-(e)]
  and the bias voltage [(f)-(h)].
  The dashed lines present the first-order contributions.
  The parameters are the same as in Fig.~\ref{Fig:1}.}
\end{figure*}

In Fig.~\ref{Fig:1}(c)-(h) we also showed the TMR calculated in
the first order (sequential) approximation (dashed lines). It is
evident that the role of cotunneling processes is particularly
pronounced in the blockade regions, where the cotunneling
processes dominate over the sequential ones, and lead to a
significant enhancement (or reduction) of TMR. Outside the
blockade regions, TMR is determined mainly by sequential
transport, so the contribution from cotunneling processes is
rather minor.

{\it Results on shot noise:} Upon calculating the current $I$ and
the zero-frequency current noise $S$, one can determine the Fano
factor $F$, $F=S/(2e|I|)$. The Fano factor describes the deviation
of $S$ from the Poissonian shot noise given by $S_p=2e|I|$. The
bias and gate voltage dependence of the Fano factor in the
parallel and antiparallel magnetic configurations is shown in
Fig.~\ref{Fig:2}(a) and (b), respectively. When $|eV|\lesssim
k_{\rm B}T$, $S$ is dominated by thermal Nyquist-Johnson noise,
otherwise the noise due to the discrete nature of charge (shot
noise) dominates. \cite{blanterPR00} In the limit of $V\rightarrow
0$, the current tends to zero whereas the current noise is
dominated by thermal noise. This leads to a divergency of the Fano
factor in the linear response regime. Therefore, in the density
plots of $F$ we mark the low bias voltage regime with a black
line.

In the cotunneling regime, where the dot is empty (or fully
occupied), the shot noise is Poissonian, Fig.~\ref{Fig:2}(c). This
is because the current flows then only due to elastic
non-spin-flip second-order processes.
\cite{thielmann,sukhorukovPRB01} However, in the Coulomb blockade
regions, where inelastic cotunneling processes also contribute, we
find a pronounced super-Poissonian shot noise, see Fig.
~\ref{Fig:2}(d) and (e). This increased shot noise is related to
bunching of electrons carried by different types of cotunneling
processes. Furthermore, in the case of magnetic leads there is an
additional contribution to the noise coming from the difference
between the spin-up and spin-down channels which leads to bunching
of fast cotunneling processes involving the majority electrons.
This is more pronounced in the parallel configuration where the
difference between the two channels is approximately equal to
$(1+p)^2/(1-p)^2$, while for the antiparallel configuration the
two channels are comparable, see Fig.~\ref{Fig:2}(a) and (b). The
variation of the Fano factor with the bias voltage is displayed in
Fig.~\ref{Fig:2}(f)-(h). When the dot is empty
[Fig.~\ref{Fig:2}(f)], the Fano factor is Poissonian and starts to
drop at the onset of sequential tunneling. With increasing bias
voltage, the noise becomes super-Poissonian in some range of the
transport voltage. In the case of singly and doubly occupied
quantum dot, the bias voltage dependencies of $F$ are
qualitatively similar. In the Coulomb blockade regime the current
noise becomes super-Poissonian (up to $F\approx 2.5$ in the
parallel configuration), indicating the existence and role of
inelastic and spin-flip cotunneling processes.
\cite{sukhorukovPRB01} For example, in the case of doubly occupied
dot, Fig.~\ref{Fig:2}(h), the maximum Fano factor is found for
voltages where inelastic cotunneling between the two orbital
levels is allowed, $|eV| \approx 2\delta\varepsilon$. On the other
hand, in the sequential tunneling regime the Fano factor becomes
suppressed and is generally sub-Poissonian due to the Coulomb
correlations in sequential transport. \cite{braun04,weymannJPCM07}
For comparison, in Fig.~\ref{Fig:2}(c)-(h) we show the results
calculated within the sequential tunneling approximation (dashed
lines), which clearly show the role of cotunneling processes in
shot noise, particularly in the blockade regions. We also note
that the occurrence of super-Poissonian shot noise in the Coulomb
blockade regime has also been reported experimentally in quantum
dots coupled to nonmagnetic leads.
\cite{gossardPRB06,onacPRL06,gossardPRL07}

In conclusion, we have discussed the spin-polarized transport
through a two-level quantum dot coupled to ferromagnetic leads in
the sequential and cotunneling regimes. We have analyzed the
dependence of the TMR and Fano factor on the bias and gate
voltages. In the Coulomb blockade regime we have found a
suppression of TMR for singly and doubly occupied quantum dot.
Furthermore, in these transport regimes the inelastic cotunneling
processes lead to super-Poissonian shot noise, irrespective of
magnetic configuration of the system. On the other hand, in the
sequential tunneling regime, the current noise is generally
sub-Poissonian, while TMR exhibits oscillatory-like dependence on
the bias voltage. We also notice that generally the sequential
tunneling approximation in the Coulomb blockade regime
underestimates the TMR and overestimates the Fano factor.

We acknowledge discussions with A. Thielmann. This project was
supported by funds of the Polish Ministry of Science and Higher
Education as a research project in years 2006-2009. I.W.
acknowledges support from the Foundation for Polish Science.


\end{document}